%% file: paper.tex
\documentclass[conference]{IEEEtran}

\usepackage{amsmath}
\usepackage{graphicx}
\usepackage{epstopdf} 
\usepackage{multirow}

\graphicspath{{./figures/}}

\usepackage{booktabs}

\usepackage[justification=centering, labelfont=bf]{caption}
\usepackage{subcaption}
\usepackage{listings}
\usepackage{verbatim}
\usepackage{url}
\usepackage{footnote}
\usepackage{array}
\usepackage{hhline}
\usepackage{tikz}
\usepackage{rotating}
\usepackage{amsfonts}
\usepackage{algorithm}
\usepackage[noend]{algpseudocode}
\usepackage{tikz}
\usepackage{graphicx,pifont}
\usepackage{adjustbox}
\usepackage{breqn}
\usepackage{changepage,threeparttable} 
\usepackage{tablefootnote}

\makeatletter
\def\BState{\State\hskip-\ALG@thistlm}
\makeatother

\let\origthelstnumber\thelstnumber
\makeatletter
\newcommand*\Suppressnumber{%
  \lst@AddToHook{OnNewLine}{%
    \let\thelstnumber\relax%
     \advance\c@lstnumber-\@ne\relax%
    }%
}

\newcommand*\Reactivatenumber[1]{%
  \setcounter{lstnumber}{\numexpr#1-1\relax}
  \lst@AddToHook{OnNewLine}{%
   \let\thelstnumber\origthelstnumber%
   \refstepcounter{lstnumber}
  }%
}

\newcommand{\helex}{HeLEx}
\newcommand{\dfgs}{12}
\newcommand{\cgras}{9}
\newcommand{\reductionOps}{68.7\%}
\newcommand{\reductionOpsHeat}{43.6\%}
\newcommand{\reductionOpsBB}{56.4\%}
\newcommand{\reductionCommonOps}{46.5\%}
\newcommand{\reductionArea}{70\%}
\newcommand{\reductionPower}{51\%}
\newcommand{\overMinimum}{6.2\%}
\newcommand{\speedupOverExisting}{2.6X}

\begin{document}

\title{\helex: A Heterogeneous Layout Explorer for Spatial 
Elastic Coarse-Grained Reconfigurable Arrays}

\author{\IEEEauthorblockN{Alan Jia Bao D and Tarek S.\ Abdelrahman}
\IEEEauthorblockA{ University of Toronto\\
Email: alan.du@mail.utoronto.ca, tsa@ece.utoronto.ca}}
 
\maketitle

\input{abstract}

\renewcommand{\algorithmicrequire}{\textbf{In:}}
\renewcommand{\algorithmicensure}{\textbf{Out:}}
\newcommand{\In}{\Require}
\newcommand{\Out}{\Ensure}

\input{intro.tex}

\input{background.tex}

\input{helex.tex}

\input{evaluation.tex}

\input{related.tex}

\input{conc.tex}
\bibliographystyle{IEEEtran}
\bibliography{paperbib}

\end{document}

%% file: abstract.tex
\begin{abstract}

We present \helex, a framework for determining the functional 
layout of heterogeneous spatially-configured elastic Coarse-Grained
Reconfigurable Arrays (CGRAs). Given a collection of input 
data flow graphs (DFGs) and a target CGRA, the framework starts 
with a full layout in which every processing element (PE) supports 
every operation in the DFGs. It then employs a branch-and-bound (BB) 
search to eliminate operations out of PEs, ensuring that the 
input DFGs successfully map onto the resulting CGRAs, eventually 
returning an optimized heterogeneous CGRA. Experimental evaluation
with 12 DFGs and 9 target CGRA sizes reveals that the framework
reduces the number of operations by \reductionOps\ on average, resulting
in a reduction of CGRA area by almost \reductionArea\ and of power 
by over \reductionPower, all compared to the initial full layout. 
\helex\ generates CGRAs that are on average only within \overMinimum\
of theoretically minimum CGRAs that support exactly the number of 
operations needed by the input DFGs. A comparison with functional 
layouts produced by two state-of-the-art frameworks indicates that 
\helex\ achieves better reduction in the number of operations, by 
up to \speedupOverExisting.

~
\end{abstract}

%% file: intro.tex
\section{Introduction}
\label{sec:intro}

Coarse-Grained Reconfigurable Arrays~(CGRAs) are receiving increased interest, 
particularly in the domain of high-performance computing~\cite{cgra4hpc22},
mainly for their ability to re-configure while providing area and power 
efficiency close to those of ASICs~\cite{Liu2019,Podobas20survey}. A CGRA is 
typically a two-dimensional grid of processing elements (PEs) and programmable 
links. The operations performed by the PEs and the connectivity of the links 
are configured by software. A CGRA is utilized by {\em mapping} a data flow 
graph (DFG)---a directed acyclic graph with nodes  representing operations and 
edges representing the flow of data between operations---to the PE grid. Each 
DFG node is assigned to a PE, which is configured to perform the operation of 
the node, and the links are configured to reflect inter-node 
connectivity. Instances of the DFG are then executed in a pipelined fashion for 
high throughput.

CGRAs are typically designed to be {\em homogeneous}, i.e., with every PE 
supporting the same set of operations. However, this homogeneity leads to 
area and power inefficiencies~\cite{revamp, heta, hiercgra}. A {\em heterogeneous} 
design, in which each PE supports only a subset of the operations, improves area 
and power efficiency, but poses two key challenges: (1) how to determine what operations 
are to be supported by each PE, and (2) how to ensure DFG mapping success to the
heterogeneous CGRA. In other words, how to determine the {\em functional layout} 
of the CGRA while guaranteeing mapping success.

In this work, we address these challenges in the context of {\em elastic spatially
configured} CGRAs. Specifically, given a set of DFGs and a target CGRA, we determine 
a heterogeneous functional layout that minimizes area and power consumption, while 
ensuring mapping success across the given set of DFGs.

We propose, implement and evaluate the \underline{He}terogeneous \underline{L}ayout 
\underline{Ex}plorer (\helex) framework. The framework starts with a full 
homogeneous CGRA where each PE supports every operation. It then employs a 
branch-and-bound (BB) search~\cite{Morrison2016} to progressively remove support 
for some operations from individual PEs (hereafter expressed as removing operations 
from PEs for brevity). This is done while ensuring that all input DFGs can 
still successfully map to the resulting CGRAs. It produces as output a 
heterogeneous CGRA that minimizes area and power consumption.
The BB search is guided by a cost function that is based on estimates of
the areas of the PE components obtained using the Synopsys Design Compiler 
(DC)~\cite{synopsys}.

Our evaluation of \helex\ using \dfgs\ DFGs and \cgras\ target CGRAs 
shows that the framework reduces the number of operations of the PEs by 
\reductionOps\ on average, without sacrificing the mapping success of the input 
DFGs. This reduction leads to heterogeneous CGRAs that have \reductionArea\ less 
area and consume \reductionPower\ less power, compared to the full homogeneous ones. 
Indeed, \helex's generated CGRAs are on average only within \overMinimum\ of ones that 
contain the minimal number of operations needed by the input DFGs.
Comparison with functional layouts obtained by two state-of-the-art frameworks 
indicates that \helex\ achieves better reduction in the number of operations, 
by up to \speedupOverExisting.

Thus, the contributions of this work are:
\begin{itemize}
\item A novel framework for generating heterogeneous functional layouts for
spatially-configured elastic CGRAs. 
\item An evaluation that shows that significant area/power reductions can
be achieved through the framework.
\end{itemize}

The remainder of this paper is organized as follows. Section~\ref{sec:background} 
gives background material. Section~\ref{sec:helex} details the \helex\ framework.
Section~\ref{sec:evaluation} presents our evaluation. Section~\ref{sec:related} 
reviews related work. Finally, Section~\ref{sec:conclusion} concludes.

%% file: background.tex
\section{Background}
\label{sec:background}

\subsection{CGRAs}

A CGRA is typically a two-dimensional grid of processing elements (PEs)
interconnected by programmable links. The design space of CGRAs has
been explored in terms of the complexity and interconnection of 
PEs~\cite{Mei03,chen18, Govindaraju12, Kasgen21}, the configuration 
approach~\cite{Podobas20, Nowatzki17, Lee09}, the execution 
model~\cite {Mei03, Nowatzki17, Govindaraju12, Ragheb22, Podobas20}, 
and how control flow is handled~\cite{Chin18, Han13, Karunaratne19, 
Balasubramanian18}.

\begin{figure}[t]
  \centering
  \includegraphics[width=0.90\linewidth]{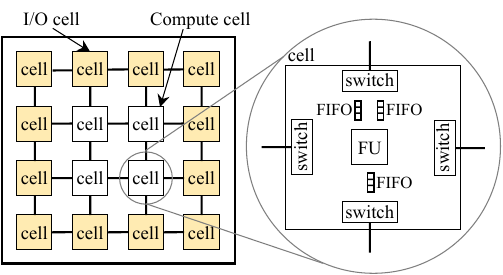}
  \caption{Target CGRA}
  \label{fig:tcgra}
\end{figure}

We target the T-CGRA architecture depicted in
Fig.~\ref{fig:tcgra}~\cite{Abbaszadeh23}. It consists of an array of PEs, 
which we refer to as {\em cells}. The cells are interconnected in a 
4-nearest-neighbor (4NN) topology. Each cell contains a functional unit (FU),
which has an ALU that performs integer and floating-point arithmetic, 
as well as logical operations. Each cell also contains a set of programmable 
switches that allow the output of one ALU to connect to the inputs of any 
other ALU, possibly through other cells. 
Further, each cell contains FIFOs to support an {\em elastic dynamic 
dataflow} execution model~\cite{Podobas20, Ragheb22}. 

T-CGRA is {\em spatially-configured}, i.e., the operation performed
by a cell is fixed for the duration of execution~\cite{Podobas20, 
Nowatzki17}. There are two types of cells: {\em I/O cells} on the border 
of the array that are limited to executing load and store operations, and 
{\em compute cells} in the interior of the array that execute arithmetic/logic 
operations. I/O cells contain only FIFOs and no compute elements. Thus, {\em 
this work focuses on removing functionality only from compute cells}.

T-CGRA utilizes a reserve-on-demand heuristic-based spatial mapper called 
RodMap~\cite{rodmap}. The mapper is both fast and has a high mapping success 
rate, close to 90\%. It achieves this high success rate by identifying the 
congestion that arises when more than one DFG edge is assigned to the same 
CGRA link. Based on congestion patterns, RodMap ``reserves'' CGRA cells around 
the congestion solely for routing, thus effectively increasing routing resources 
and eliminating the congestion. In this work, we use this mapper as a black box 
for mapping DFGs to both homogeneous and heterogeneous CGRAs.

\subsection{Branch-and-Bound}

Branch-and-Bound (BB) is a widely-used method for solving combinatorial 
optimization problems~\cite{Morrison2016}. 
The method {\em implicitly} enumerates all possible solutions to a problem 
by storing partial solutions called {\em subproblems} in a tree structure. 
A {\em cost} is associated with each node in the tree, which reflects the
``goodness'' of the partial solution or the subproblem. Unexplored nodes 
in the tree generate children by {\em branching}, i.e., further partitioning 
of a subproblem into smaller ones (i.e., more complete partial solutions). 
Subproblems are {\em pruned}, i.e., excluded 
from further consideration when it is determined that the cost of the best 
solution that can be obtained by branching from them is worse than that of 
a known solution called the {\em bound}. The order in which subproblems are 
examined is referred to as the {\em search strategy}. The manner in which 
a subproblem is partitioned into smaller ones is referred to as the 
{\em branching strategy}. The cost, search, and branching strategies 
are often customized for the problem being tackled. Thus, BB is more 
of a framework than a specific algorithm.

%% file: helex.tex
\section{H\lowercase{e}LE\lowercase{x}}
\label{sec:helex}

\subsection{Overview}

\helex\ takes as input a set of DFGs and a target CGRA size. It
produces as output a heterogeneous CGRA functional layout, for
the target size, that minimizes area and power consumption.
\helex\ is centered around the following key idea. Starting with an initial 
valid layout (i.e., one to which all the input DFGs successfully map onto), 
\helex\ optimizes this layout by iteratively pruning combinations of 
compute resources from the cells of the layout. In each iteration, the 
mapper is used to re-map the DFGs, forcing DFG nodes impacted by the 
removal of a resource to place on other cells that have this resource. 
This consolidates the resources used by the DFGs on the CGRA. As more 
resources are removed, the mappings of different DFGs overlaps with each 
other and share the same set of cells, resulting in more potential 
resources that can be removed. The iterative pruning process ends 
either when no more layouts that successfully map the DFGs can be found 
or a pre-defined iteration limit is reached.

\helex\ consists of three main phases: (1) determining an initial 
layout that serves as the start of the BB search, (2) a BB search
that prioritizes the elimination of more expensive and less frequently-used 
resources, and (3) a second BB search that attempts to remove 
all combinations of resources.

\begin{algorithm}
    \input{algorithms/helex-overview}
    \caption{\helex}
    \label{alg:helex-overview}
\end{algorithm}

Algorithm~\ref{alg:helex-overview} gives a high-level overview
of \helex. It takes as input a set of DFGs, the size of the target
CGRA $R \times C$,  an operation grouping (Section~\ref{sec:groups}) 
$opGroups$, the relative costs of the target CGRA components
(Section~\ref{sec:costs}) $cellCosts$, and a limit on the number
of times the mapper can be invoked $L_{test}$, which is used to 
limit search time.  The minimum 
number of operations that are needed to support the input DFGs is 
determined (Section~\ref{sec:minops}) on line 1. The initial layout 
from which the search begins is then computed (Section~\ref{sec:initial}) 
on lines 2 to 4, before the two phases of BB search are conducted 
(Section~\ref{sec:formulation}) on lines 5 and 6. 
The best layout, $best$, is returned on line 7. 

\subsection{Operation Grouping}
\label{sec:groups}

DFGs that stem from real-world applications contain  a mix of integer, 
floating point (FP), and multiplication operations, as well as less 
common operations like division, square root, logarithm, and exponents~\cite{rodmap}. 
A full (homogeneous) layout of a CGRA has each cell supporting all the 
operations (both common and less common) that appear in a given set of 
input DFGs.

In order to make the search more efficient and more accurately
reflect hardware savings, \helex\ groups individual operations into 
{\em operation groups} based on their hardware implementations. 
For example, an ALU that supports an ADD operation easily supports a SUB 
operation with minimal extra cost. In contrast, an ADD and DIV require 
different hardware, and are placed in different groups. 

We determine operation groups based on their implementation by the Synopsys 
DesignWare Library~\cite{designware}. Specifically, we group DFG operations 
into the 6 operation groups shown in 
Table~\ref{tab:opgroup}. Integer and floating point multiplies 
(divides) are grouped together because it is possible to combine them 
efficiently using the DesignWare Library.  Complex operations that 
do appear in HPC applications but are less common (i.e., exp, log, 
sqrt) are grouped together since they can be efficiently implemented using 
a multi-function unit from the  DesignWare Library, or approximated using 
table lookup (e.g.,~\cite{Suganth09}). 
Thus, \helex\ removes one operation group at a time to more accurately 
reflect hardware savings. 

\begin{table}[ht]
    \centering
    \input{tables/opgroup}
    \caption{Operation groupings used in \helex}
    \label{tab:opgroup}
\end{table}

The use of operation groups has the advantage of reducing the search 
space. Instead of searching among all operation combinations, the 
search is done among all operation group combinations. However, it 
should be emphasized that \helex's search framework is independent of a 
given operation grouping and can indeed work with any grouping.
Operation groupings can be modified to reflect different hardware 
realizations and attributes of input DFGs.

\subsection{Component Costs}
\label{sec:costs}

The BB search's cost function (Section~\ref{sec:formulation}) is 
based on the costs of individual CGRA components. These components 
implement DFG operation groups, FIFOs, switches, empty functional 
units (FUs), and empty CGRA cells. All operations are 32-bits 
wide, and floating point operations are in IEEE 754 standard. 
The costs are determined by instantiating these components using SystemVerilog and 
then synthesizing them using Synopsys DC~\cite{designware}. This synthesis
provides area estimates for each CGRA component, which are then normalized with 
respect to that of the integer arithmetic ALU. 

\subsection{Minimum Number of Group Instances}
\label{sec:minops}

The target CGRA is spatial. Thus, given a DFG and an
operation group, $g$, CGRA cells must contain at least as many 
instances of $g$ as there are operations in the DFG that require $g$.
For a set of input DFGs, the cells
must contain at least as many instances of $g$ as the maximum 
number of corresponding operations across the DFGs. 
There is no guarantee that the input DFGs can successfully map 
onto a layout with this number of instances.
This is simply a theoretical minimum on the number of instances, 
which is used to prune layouts during the search. 

\subsection{Initial Layout}
\label{sec:initial}

\begin{figure*}
    \centering
    \includegraphics[width=0.99\linewidth]{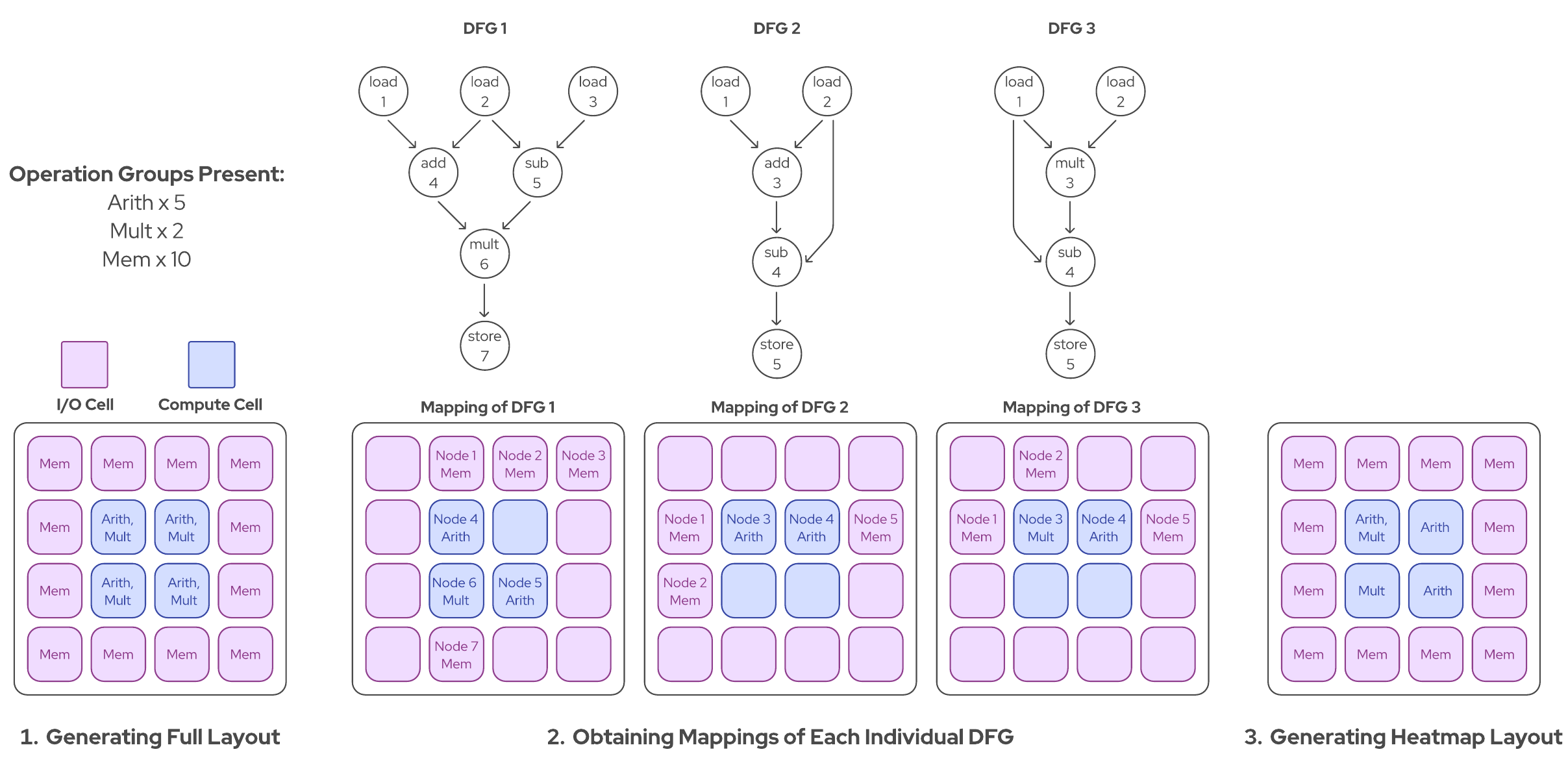}
    \vspace*{-0.15in}
    \caption{Example of generating heatmap layout from individual DFGs}
    \label{fig:heatmapprocess}
\end{figure*}

An initial layout is needed before \helex\ begins its BB search.
This is done in the following steps and is illustrated 
with the example in Fig.~\ref{fig:heatmapprocess}. First, a 
homogeneous (full) CGRA layout is generated using all of the 
operation groups present in the input DFGs 
(Fig.~\ref{fig:heatmapprocess}.1). The DFGs are then 
mapped on the full layout one by one (Fig.~\ref{fig:heatmapprocess}.2). 
Should one or more DFGs fail to map to this full layout, \helex\ 
terminates in failure. 
Otherwise, the resulting individual assignments of DFG nodes to 
CGRA cells are overlaid to create a heterogeneous \textit{heatmap} 
layout for the CGRA (Fig.~\ref{fig:heatmapprocess}.3). \helex\ only targets 
the \textit{compute cells} of a CGRA. Thus, I/O cells are 
left untouched in the heatmap layout.  

The change in a target CGRA layout affects mapping and there is no 
guarantee that all the DFGs can successfully map to the CGRA with the 
heatmap layout. Thus, the input DFGs must be re-mapped onto the heatmap 
layout. If all the DFGs successfully re-map, the heatmap layout is used 
as the initial layout. Otherwise, if one or more DFGs fail re-mapping, 
the initial layout remains the full layout. 

\subsection{The BB Search}
\label{sec:formulation}

We formulate our problem as a BB search. Each subproblem 
represents a possible functional layout of the target CGRA. 
Associated with each subproblem is a cost that is determined based 
on the costs of the components of the CGRA of the corresponding 
layout. Specifically, this cost is the sum of the costs of these components: 
\begin{equation}
\label{eq:cost}
\begin{split}
Layout\: Cost =\  & N_t \times ( cost(empty\: cells) + cost(FIFOs) ) \\
& + \sum\limits_{g} N_g \times cost(g) \\
\end{split}
\end{equation}
where $N_t$ is the total number of compute cells, 
and $N_g$ is the number of instances of each group $g$. The costs of the
components is determined as described in Section~\ref{sec:costs}.
The goal is to find a layout with the smallest cost that successfully maps 
the input DFGs.

The branching strategy generates new subproblems from a given one by 
removing one or more operations from a cell. Two branching strategies 
are used in \helex. The first, referred
to as {\em operation-based subproblem generation} (OPSG), limits the
removal of operations to one group at a time, from highest to lowest cost. 
The second considers all possible combinations of operation
removals in no particular order. It is referred to as {\em general 
subproblem generation} (GSG). Both are explained below.

The bound of the BB search starts as the cost of the initial 
layout. If a subproblem is generated
that has a lower cost than the bound, the subproblem is
tested for feasibility by mapping the input DFGs onto the CGRA with
a layout corresponding to this subproblem. If all the DFGs successfully
map, the bound is updated with the cost of this subproblem, which
becomes the current best layout. 

A {\em best-first} search strategy is employed, where the subproblem
with the smallest cost is considered first. Therefore, subproblems are
inserted on a min-priority queue. Pruning is realized using the minimum 
number of group instances, $minInsts$. A layout that contains a number of group 
instances less than the minimum number of instances determined by $minInsts$ 
cannot possibly be successful in mapping and is eliminated.
Further, since the search can make multiple identical attempts at 
removing the same operation group from a cell, a subproblem is also pruned if
the same group is removed multiple times from the cell, each time
resulting in mapping failure.

\subsubsection{Operation Based Subproblem Generation (OPSG)}

\helex\ begins the search by restricting the potential subproblems that 
can be generated to focus on only a single group at a time, from highest 
to lowest cost. The goal is to prioritize the removal of 
more expensive operations in a systematic approach and verify 
the feasibility of the resulting layouts by {\em selectively testing}
the layouts. 

The pseudocode of the BB search with OPSG is shown in Algorithm~\ref{alg:helex-opsg}. 
It takes as input an initial layout of size R~$\times$~C, a set of DFGs, the 
minimum number of operation group instances, {\em minGroups}, and a limit on 
how many layouts can be tested before the search terminates, $L_{test}$.

\begin{algorithm}[t]
    \input{algorithms/helex-opsg}
    \caption{BB Search with OPSG}
    \label{alg:helex-opsg}
\end{algorithm}

The operation groups, {\em opGroups}, are sorted in descending order based on
their costs, {\em cellCosts}. They are inserted in this order on the list
{\em removalOrder} (line 2). The best layout is set as the initial 
layout (line 1) and the counter that tracks the number of tested layouts is
initialized to 0 (line 3).

Iterating over \textit{removalOrder} (line 4), a BB search is conducted 
for the current operation group, \textit{opType}. A min-priority queue, 
\textit{pq}, is initialized on line 7 and filled with all possible valid 
layouts generated from the current best layout, \textit{bestLayout}, 
on line 8. A new layout is generated by removing \textit{one} instance 
of \textit{opType} from a single cell on the original layout, which 
in this case is \textit{bestLayout}. All possible combinations of removing 
\textit{one} instance of \textit{opType} from a single cell, starting from 
the top-left cell down to the bottom-right cell, are generated and pushed 
into the queue if valid, i.e. meet the \textit{minGroups} requirement. 

While \textit{pq} is not empty, $L_{test}$ has not been reached, and a new 
best layout has not been found (line 10), layouts are popped from the queue 
and tested for feasibility if their cost is lower than the cost of 
\textit{bestLayout} (lines 11-14). The \textit{numTested} counter is 
incremented (line 13) and if testing is successful, the best layout is 
updated to the current one (line 15). The \textit{newBestFound} flag is 
also set (line 16) to exit the inner {\tt While} loop. Since each layout was 
generated by removing the same operation group from a different cell in the 
same original layout, the cost of all layouts in the queue are the same. 
Thus, it is not necessary to look beyond the first successful layout that 
improves upon the best layout. Further, because the difference between the 
original layout, which is known to be feasible, and the current layout is 
a single operation group, it is possible to \textit{selectively test} these 
layouts (line 14). This means only needing to test a layout with the DFGs 
that contain the operations that were removed rather than the entire set of 
DFGs. However, if the original layout is not known to be feasible and the 
difference between the current and original layouts is too great, selective 
testing can no longer be utilized. 

After exiting the inner {\tt While} loop, the queue must be re-initialized 
using the updated best layout and repeated until no more valid layouts can 
be generated for the given \textit{opType}, $L_{test}$ has been reached, all 
generated layouts fail testing. At this point, the \textit{stopSearchRound} 
flag is set (lines 17-18) and the outer {\tt While} loop is stopped, ending 
the BB search for the given \textit{opType}. This process repeats with all 
other operation groups, until there are no more groups remaining, or $L_{test}$ 
has been reached. Finally, the best layout and the number of layouts that 
were tested are returned. 

It is possible to selectively test a layout with only the DFGs that contain 
the operations that were removed rather than the entire set of DFGs. This
is possible because the mapping of DFGs that do not have nodes with these 
operations is unaffected by their removal.
If testing is successful, the best layout is updated to the current one 
(line 15), and the $newBestFound$ flag is set (line 16) to exit the inner 
{\tt While} loop. This is because the costs of all subproblems in the 
queue are the same.  Thus once the first successful layout that improves 
upon the best layout is found, the queue must be re-initialized using the 
updated best layout.  The process repeats with other operation groups until 
no better layouts can be found, or the $L_{test}$ is reached (line 17-18); 
$best$ and $nMapped$ are returned.

\subsubsection{General Subproblem Generation (GSG)}

\helex\ removes additional operation groups by allowing
the removal of any operation group in a cell in no particular 
order. The goal is to explore all possible layout 
combinations.

\begin{algorithm}
    \input{algorithms/helex-failchart}
    \caption{BB Search with GSG}
    \label{alg:helex-gsg}
\end{algorithm}

The pseudocode of the BB search with GSG is shown in Algorithm \ref{alg:helex-gsg}. 
It takes in the same inputs as the OPSG algorithm and also the number of layouts 
that have been tested by OPSG, \textit{numTested}. The starting layout for the 
GSG algorithm is the best layout determined by the OPSG phase.

The GSG algorithm is similar to the OPSG one (Algorithm~\ref{alg:helex-opsg}), 
with some key differences.  First, in this phase, a new layout is generated 
by removing \textit{any} combination of operation groups from a single cell 
in the current layout. \textit{All} possible combinations of removing 
\textit{any} one or more operation groups from a single cell, starting from 
the top-left cell down to the bottom-right cell, are generated and stored 
into the queue if the layout is valid (line 3). This is unlike the OPSG 
phase where only a single type of operation group is removed at a time. 
The operation groups removed to generate the new layout from the current 
layout is stored as metadata for future use. 

Second, the {\tt While} loop of this phase does not terminate after finding 
the first successful layout that improves upon the best solution. Rather, 
it stops when no more valid layouts remain in \textit{pq} or when $L_{test}$ 
has been reached (line 5). 

Lastly, a new layout can no longer be selectively tested with only the DFGs 
that contain the operation group that was removed to obtain the layout being 
tested. Instead, a new layout must be tested across the entire set of DFGs 
(line 9). Unlike the OPSG phase where the difference between the current 
layout and the best layout is always a single operation group, the queue in 
the GSG phase contains layouts generated from different points in the search. 
Tracking the modification history of each layout and having confidence in their 
feasibility by only using a subset of DFGs becomes a difficult task given the 
number of layouts explored by \helex\ during the search. 

In generating all possible combinations of operation group removals, it is 
possible for two generated layouts to remove the same combination of 
operation groups from the same cell. If this combination of removal and
cell fails multiple times, chances are it will fail again. This is kept
track of in a structure called \textit{failChart} (line 4). It is 
updated when a layout has failed (line 15), reset when a successful 
better layout has been found (line 12), and used to prune subproblems 
that fail multiple times (lines 8-10).

There are other optimizations that are not reflected in 
Algorithm~\ref{alg:helex-gsg} for simplicity. We conduct the GSG BB search 
described above twice, and also prune the priority queue of subproblems that 
are too far away in cost from the best layout after failing to improve $best$ 
for more than a user-specified number of consecutive iterations.

%% file: algorithms/helex-overview.tex
\begin{algorithmic}[1] 
    \Require DFGs, R, C, opGroups, cellCosts, $L_{test}$
    \Ensure bestLayout
    \State minGroups $\gets$ findMinGroups(DFGs)
    \State initialLayout $\gets$ createHeatmapLayout(DFGs, R, C, opGroups)
    \If{testLayout(initialLayout, DFGs, R, C) \textbf{is not} successful}
        \State initialLayout $\gets$ createFullLayout(R, C, opGroups)
    \EndIf
    \State bestLayout, numTested $\gets$ runOpsgBB(\parbox[t]{.3\linewidth}{initialLayout, DFGs, R, C, opGroups, cellCosts, minGroups, $L_{test}$)}
    \State bestLayout $\gets$ runGsgBB(\parbox[t]{.5\linewidth}{bestLayout, DFGs, R, C, opGroups, cellCosts, minGroups, numTested, $L_{test}$)}
    \State \textbf{return} bestLayout
\end{algorithmic}

%% file: tables/opgroup.tex
\begin{tabular}{ll}\toprule
\textbf{Group} &\textbf{Description} \\\midrule
Arith &Integer and logic ops (excluding DIV and MULT) \\
Div &Integer and floating point DIV \\
FP &Floating point ops (excluding DIV and MULT) \\
Mem &Memory ops (LOAD, STORE) \\
Mult &Integer and floating point MULT \\
Other &Special ops (EXP, LOG, SQRT, etc.) \\
\bottomrule
\end{tabular}

%% file: algorithms/helex-opsg.tex
\begin{algorithmic}[1] 
    \Require initialLayout, DFGs, R, C, opGroups, cellCosts, minGroups, $L_{test}$
    \Ensure bestLayout, numTested
    \State bestLayout $\gets$ initialLayout
    \State removalOrder $\gets$ sort(opGroups, cellCosts)
    \State numTested $\gets$ 0
    \For{opType \textbf{in} removalOrder}
        \State stopSearchRound $\gets$ False
        \While{stopSearchRound \textbf{is} False}
            \State pq $\gets$ initPriorityQueue()
            \State pq $\gets$ generateValidOPSGLayouts(bestLayout, opType, minGroups)
            \State newBestFound $\gets$ False
            \While{pq.length $>$ 0 \textbf{and} numTested $< L_{test}$ \textbf{and} newBestFound \textbf{is} False}
                \State currentLayout $\gets$ pq.pop()
                \If{currentLayout.cost $<$ bestLayout.cost}
                    \State numTested++
                    \If{selectiveTestLayout(currentLayout, DFGs, R, C, opGroups)}
                        \State bestLayout $\gets$ currentLayout
                        \State newBestFound $\gets$ True
                    \EndIf
                \EndIf
            \EndWhile
            \If{newBestFound \textbf{is} False}
                \State stopSearchRound $\gets$ True
            \EndIf
        \EndWhile
    \EndFor
    \State \textbf{return} bestLayout, numTested
\end{algorithmic}

%% file: algorithms/helex-failchart.tex
\begin{algorithmic}[1] 
    \Require initialLayout, DFGs, R, C, opGroups, cellCosts, minGroups, numTested, $L_{test}$, $L_{fail}$
    \State bestLayout $\gets$ initialLayout
    \State pq $\gets$ initPriorityQueue()
    \State pq $\gets$ generateValidGSGLayouts(bestLayout, minGroups)
    \State failChart $\gets$ initFailChart(R, C)
    \While{pq.length $>$ 0 \textbf{and} numTested $<$ $L_{test}$}
        \State currentLayout $\gets$ pq.pop() 
        \If{currentLayout.cost $<$ bestLayout.cost}
            \If{failChart[(currentLayout.removedOp, currentLayout.cell)] $<\ L_{fail}$}
            \State succ $\gets$ testLayout(currentLayout, DFGs, R, C)
            \State numTested++
            \EndIf
            \If{succ}
                \State failChart $\gets$ initFailChart(R, C)
                \State bestLayout $\gets$ currentLayout
            \Else
                \State failChart[(currentLayout.removedOp, currentLayout.cell)]++
                \State \textbf{continue}
            \EndIf
        \EndIf
        \State pq $\gets$ expandSubproblems(currentLayout, minGroups, failChart, $L_{fail}$)
    \EndWhile

    \State \textbf{return} bestLayout
\end{algorithmic}

%% file: evaluation.tex
\section{Evaluation}
\label{sec:evaluation}

\helex\ is implemented and evaluated on an Ubuntu 
22.04 workstation with an Intel i9-13900K processor (3.7~GHz) and 
64~GBs of DRAM. 
Table~\ref{tab:dfgs-used} shows the DFGs used in the evaluation, giving 
for each the number of nodes, $V$, the number of edges, $E$, and a brief 
description. The DFGs are mapped using our target CGRA's mapper~\cite{rodmap}, 
with its default parameters. The DFGs are 
mapped to 9 CGRA instances, ranging in size from $10 \times 10$ to 
$13 \times 15$. All DFGs successfully map to the full homogeneous CGRA 
instances with these sizes. The $10 \times 10$ is the smallest size onto
which {\em all} our DFGs successfully map. The CGRA instance sizes we use
represent large CGRA sizes (as opposed to the typical $4 \times 4$ or 
$6 \times 6$ sizes~\cite{revamp, heta,Karunaratne19}), demonstrating the 
scalability of \helex.

\begin{table}[t]
    \centering
    \input{tables/dfgs-used}
    \caption{DFG benchmarks used for evaluation}
    \label{tab:dfgs-used}
\end{table}

The operations of the DFGs are grouped into the 6 operation 
groups shown earlier in Table~\ref{tab:opgroup}. The costs associated with 
the CGRA components that implement these groups and other CGRA components
are shown in Table~\ref{tab:cellcost}. These costs are obtained using the 
Synopsys DC~\cite{synopsys} with the 45nm FreePDK45 and Nangate Open Cell 
library~\cite{freepdk45} synthesized at $\sim$220~MHz with no timing 
violations and give the area and power costs for each component. 

We run \helex\ given the set of input DFGs and the 9 aforementioned CGRA 
instances. We utilize a large enough value of $L_{test}$ to ensure that the 
search ends only when no more sub-problems can be generated. This value is 
set to an experimentally-determined value of 2000 for a $10 \times 10$ CGRA 
instance and is increased with instance size since more compute cells require 
more iterations to prune.

We assess the quality of the heterogeneous layout that \helex\ produces by 
the reduction in the number of instances of operation groups, the reduction
in area, and the reduction in power, in comparison to the corresponding 
full homogeneous layout. Since \helex\ targets only the resources of compute 
cells, reduction is with respect to the full resources of the compute cells.

\begin{table}[t]
    \centering
    \input{tables/cellcost}
    \caption{CGRA component costs used in \helex}
    \label{tab:cellcost}
\end{table}

\subsection{Operation Group Instance Reduction}
\label{sec:opresults}

\begin{figure}[t]
  \centering
  \includegraphics[width=0.90\linewidth]{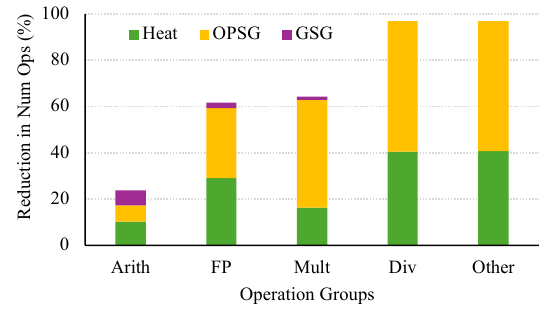}
  \vspace*{-0.05in}
  \caption{Reduction in number of operations groups}
  \vspace*{0.04in}
  \label{fig:mainresults-opbreakdown}
\end{figure}

Fig.~\ref{fig:mainresults-opbreakdown} shows the reduction in the number 
of instances of each operation group achieved by \helex, averaged over the 
target CGRA sizes. The reduction is broken down by contribution from the 
use of the heatmap, and by the BB search. Some observations can be drawn 
from the figure.

First, \helex\ achieves a significant reduction in the number of group 
instances, averaging \reductionOps. This reduction is from all operation
groups, but is more noticeable for the Div and Other operation groups.
This is expected since there is a small number of these operations in the 
input DFGs. Nonetheless, excluding these two groups still gives a notable 
average reduction of \reductionCommonOps. 

Second, the reduction in the 
number of group instances due to starting from the heatmap, and from the BB 
search is \reductionOpsHeat, and \reductionOpsBB\ respectively. This reflects 
the importance of using the heatmap when it is available and also 
demonstrates the ability for the search to contribute to the removal of 
operation group instances with or without the heatmap starting layout. 

Third, the average reduction contributed by starting at the heatmap, running 
the OPSG search, and running the GSG search is 27.4\%, 39.3\%, and 2.0\% 
respectively. This reflects that all components of \helex\ contribute to 
the removal of operation group instances. Thus, the heatmap and OPSG perform 
the bulk of reduction, accounting for 93.3\% of the total. However, GSG has
most impact in the {\tt Arith} group where it is responsible for 27.2\% of 
the total reduction. Thus, while the pruning impact of GSG is low overall,
it remains key for removing the important group of {\tt Arith} operations.

\subsection{Area and Power Reduction}
\label{sec:arearesults}

\begin{figure}[ht]
  \centering
  \includegraphics[width=0.90\linewidth]{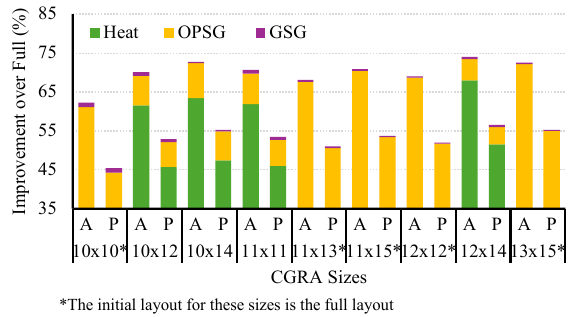}
  \caption{Improvement in area (A) and power (P)}
  \label{fig:mainresults-combined}
\end{figure}

The impact of the reduction in the number of instances of operation groups
on area and power is shown in Fig.~\ref{fig:mainresults-combined}, 
for each target CGRA size. Again, some observations can be made.

First, there is a reduction in area and power that is commensurate with the
reduction in the number of group instances. Averaged over all the CGRA sizes, 
the reduction in area is 69.4\% (peaking at 73.5\%) and in power is 52.3\% 
(peaking at 56\%). 

Second, it is not always possible to start the search from the heatmap. In 
only 4 out of the 9 target CGRAs is it possible to do so. However, when 
available, the heatmap accounts for 89\% of the total area and power reduction
obtained. This reflects the importance of our initial layout selection. 
Nonetheless the reduction due to the BB search is also noteworthy, reducing total 
area and power by 41.1\% and 31.1\% on average. Even when it is possible to 
start with the heatmap, the BB search further reduces area and power by another 
7.5\% and 6.3\% on average, demonstrating its value.

Third, there is minimal difference in area and power reduction between target 
CGRAs for which the initial layout is the heatmap and those for which the initial
layout is the full one. On average, starting with the heatmap achieves an area (power) 
reduction of 71.2\% (53.9\%) versus 68\% (51\%) when starting with the full layout.

\subsection{Search Performance}
\label{sec:searchresults}

Table \ref{tab:state-info} shows the number of subproblems expanded 
($S_{exp}$) and subproblems tested for feasibility by the mapper 
($S_{tst}$). It also shows the times (in hours) taken by the OPSG 
phase ($T_{opsg}$), the GSG phase ($T_{gsg}$) search, and the 
total \helex\ time. 

\begin{table}[ht]
    \centering
    \input{tables/searchperf}

    \vspace*{-0.1in}
    \caption{No. of subproblems and search time (hours)}
    \label{tab:state-info}
\end{table}

The average number of subproblems expanded is 1.07m and the 
average number of sub-problems tested is 2380. This reflects that
\helex\ is selecting a small fraction of them to validate with the 
mapper. On average, the ratio of 
$S_{tst}$ to $S_{exp}$ is 0.12. For layouts that start from the 
heatmap the ratio is lower at 0.005, and is 0.215 for layouts that 
start from the full layout. 

On average, the total search takes 20.8 hours to finish with OPSG 
taking up 10.8 hours and GSG taking up 10.1 hours. For layouts that 
start from the heatmap, the average search time is 13.6 hours and is
26.7 when starting from the full layout. These search times are
not unusual for design space exploration frameworks, and are indeed 
on the same order of magnitude taken by frameworks such as HETA and 
REVAMP (Section~\ref{sec:stateoftheart}).

\begin{figure}
\centering
\begin{subfigure}{0.50\textwidth}
  \centering
  \vspace*{-0.05in}
  \includegraphics[width=0.90\linewidth]{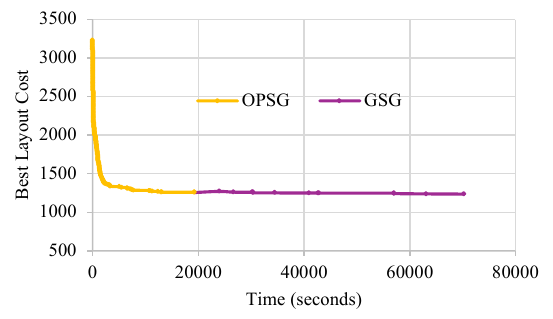}
  \vspace*{-0.10in}
  \caption{Cost of best layout over time (seconds)}
  \vspace*{0.15in}
  \label{fig:10x10-costtime}
\end{subfigure}
\hfill
\begin{subfigure}{0.50\textwidth}
  \centering
  \vspace*{-0.05in}
  \includegraphics[width=0.90\linewidth]{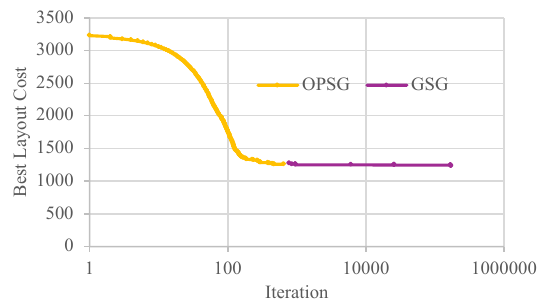}
  \vspace*{-0.10in}
  \caption{Cost of best layout over iterations}
  \label{fig:10x10-costiter} 
\end{subfigure}
\vspace*{-0.10in}
\caption{Cost of best layout over the search}
\label{fig:10x10-costsearch}
\end{figure}

Fig.~\ref{fig:10x10-costsearch} shows the rate at which \helex\ finds 
new intermediate solutions for $10 \times 10$ CGRA size. This is
representative of the results for other CGRA sizes. Specifically, the
figure shows the decrease in the cost of the best layout as a function
of the search time Fig.~\ref{fig:10x10-costtime} and the number of 
iterations Fig.~\ref{fig:10x10-costiter}. 
The figures indicate that \helex\ quickly determines 
a layout that is relatively close in cost to the final layout and spends 
the remaining time validating this by exploring other layouts and making 
incremental improvements. Indeed, \helex\ is able to land on a layout 
that is 90\% as good as the final layout in the first 30 minutes into 
the search. Thus, if search time is a primary constraint, terminating 
\helex\ early will still result in significant savings.

\begin{figure}[h!]
  \centering
  \includegraphics[width=0.90\linewidth]{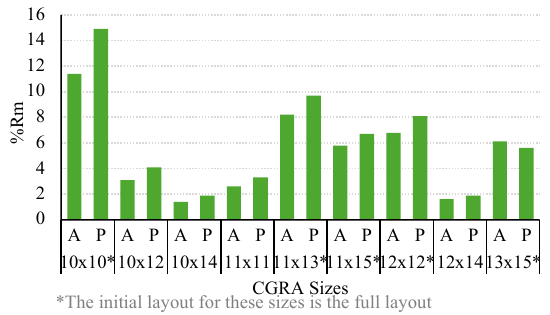}
  \vspace*{-0.00in}
  \caption{Theoretical reduction remaining}
  \label{fig:mainresults-remaining}
\end{figure}

Fig.~\ref{fig:mainresults-remaining} shows the percentage reduction
in area and power remaining to reach those of the theoretical minimum 
number of group instances (Section~\ref{sec:minops}), $\%Rm$.
On average, 94.8\% of total area and 93.8\% of total power reductions 
have been obtained, reflecting that area and power can only be further
reduced by 5.2\% and 6.2\% before the theoretical minimum is reached. 
This is even smaller at 2.2\% for area and 2.8\% for power for layouts
that start from the heatmap. This attests to \helex's ability to remove 
excess compute resources. 

\subsection{Area and Power Estimates Validation}
\label{sec:validation}

\helex\ estimates the cost of a subproblem or a layout based on the costs of 
its components (Equation~\ref{eq:cost}). We validate this cost modeling 
approach by using Synopsys DC compiler to synthesize both full and \helex's 
final optimized layouts for complete $8 \times 8$ and $12 \times 12$ CGRAs. 
This synthesis {\em includes} I/O cells for completeness. The actual area 
and power reported by Synopsys compared to those estimated by \helex\ are 
shown in Table~\ref{tab:dc-validation}. The table reflects that there is 
$\leq 1.4\%$ discrepancy in the estimates. Further, \helex's costs for each 
layout are almost identical to the improvements in area reported by Synopsys. 
This validates our approach to cost modeling.

\begin{table*}[h!]
    \centering
    \input{tables/dc-validation}
    \caption{Validation of \helex\ final layout (compute + I/O cells) using Synopsys DC}
    \vspace*{-0.2in}
    \label{tab:dc-validation}
\end{table*}

\subsection{Memory and Interconnect Resources}
\label{sec:otherresources}

\helex's BB search focuses on cell compute resources. 
Nonetheless, it can {\em posteriori} prune cells memory resources, 
in the form of FIFOs, or interconnect resources in the form of
multiplexers and switches. Specifically, resources that are 
never used in the mapping of any of the input DFGs can be eliminated 
without affecting functionality. 

Table~\ref{tab:fifos} shows the number of unused FIFOs out of the 
total number of FIFOs in the \helex\ generated layouts. It also
shows the reduction in area and power improvement over the full layout
that results from the removal of the FIFOs, $\%Impr$. The table 
shows that the removal of the FIFOs results in small but  noticeable 
savings. On average, the removal of unused FIFOs further reduces 
area and power by another 3.7\% and 7.2\%. 

Interconnect resources contribute far less to the overall cost 
compared to memory resources; $<$10\% of area and $<$5\% of power. 
This is less than the contribution of a single FIFO. Thus, there 
is less motivation to target these resources during the 
search or posteriori. 

\begin{table}[t]
    \centering
    \input{tables/fifos}
    \caption{Impact of removing excess memory resources}
    \label{tab:fifos}
\end{table}

\subsection{Impact of Target DFGs}
\label{sec:dfgimpact}

The extent to which \helex\ eliminates resources depends on input DFGs, 
in particular their sizes in relation to CGRA sizes. and the diversity of
the operations they have. Thus, we vary the input DFGs to show that 
\helex\ improves area and power across a variety of DFG inputs.

The 12 target DFGs are divided into six sets, shown in Table~\ref{tab:dfgsets}. 
Set S1 is a group of three DFGs meant to demonstrate \helex's performance on small 
DFG sets. Set S2 is a group of four DFGs that are similar in size to measure how 
DFG sizes impacts results. Set S3 is a selection of DFGs that only contain nodes 
in the {\tt Arith} and {\tt Mult} groups to show the performance of \helex\ on DFGs 
that do not contain uncommon expensive operations. Set S4 is a collection of five 
DFGs from the same domain (image processing), and evaluates \helex's capabilities 
on domain-specific DFGs. Finally, sets S5 and S6 are groups of size six and seven, 
meant to show \helex's ability on larger sets of DFGs, in contrast to set S1.

\begin{table*}[h]
    \centering
    \vspace*{0.15in}
    \input{tables/dfgsets}

    \caption{DFG sets used in evaluation}
    \vspace*{-0.1in}
    \label{tab:dfgsets}
\end{table*}

Each DFG set is tested with two CGRA sizes for a total of the 12 configurations,
as listed in Table~\ref{tab:dfgsets}. The full layout for each configuration is 
constructed with all of the operations present in the input set of DFGs. Thus, 
for example, if the DFG set does not contain any divide operations, the full 
layout will also not have any cells that support divide.

\begin{figure}[h]
  \centering
  \includegraphics[width=0.90\linewidth]{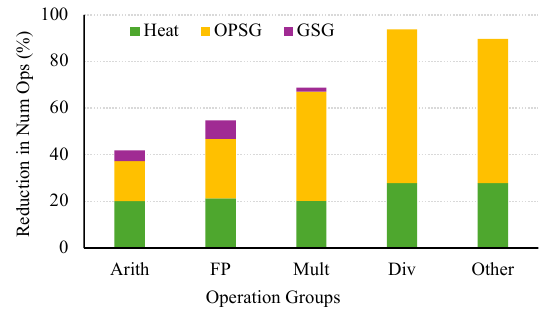}
  \vspace*{-0.05in}
  \caption{Reduction in group instances across all DFG sets}
  \label{fig:setresults-opbreakdown}
\end{figure}

Fig.~\ref{fig:setresults-opbreakdown} shows the reduction in instances of 
each operation group in the best layout compared to the full layout, averaged 
over all configurations. Similar conclusions to those in Section~\ref{sec:opresults}
can be drawn. On average, \helex\ is able to remove 69.8\% of the operation 
groups present. Excluding the uncommon {\tt Div} and {\tt Other} groups, the 
reduction is still significant, at 55.2\%. The contribution by the heatmap, 
OPSG, and GSG is 23.5\%, 43.4\%, and 2.9\% respectively. The impact of GSG is
again seen in the {\tt Arith} group, removing 55.3\% of the instances.

\begin{figure*}[h]
  \centering
  \hspace*{-0.15in}
  \includegraphics[width=0.9\linewidth]{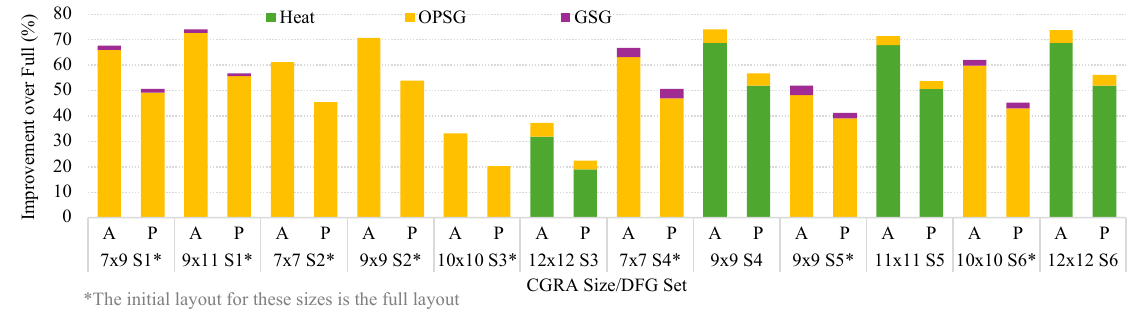}
  \vspace*{-0.05in}
  \caption{Improvement in area (A) and power (P) over full layout}
  \label{fig:setresults-combined}
\end{figure*}

Fig.~\ref{fig:setresults-combined} shows the reduction in area and power that 
the best layout achieves over the corresponding full layout of each configuration. 
Again, similar observations can be made as those in Section~\ref{sec:arearesults}. Of 
the 12 configurations, only a third are able to start from the heatmap (not marked 
by $*$). On average, across all sets, \helex\ is able to reduce area and power 
costs by 61.9\% and 46\% with respect to the full layout. The OPSG phase contributes 
to 67.7\% (67.8\%) of total area (power) savings, and the GSG contributes to 1.7\% 
(1.9\%) of total area (power) savings. The reductions are across all sets and
all configurations, which reflects that \helex\ results in area and power 
reductions for DFGs of different sizes (in relation to CGRA sizes), for 
different numbers of DFGs, and for DFGs that contain only arithmetic operations
as well as those that contain more complex operations. 

\subsection{Impact of GSG Phase}

The results in previous sections show that GSG has a small impact on reductions
in area and power, and that its impact is seen mostly in the {\tt Arith}  group. 
Indeed, if input DFGs contain only these groups and the target CGRA layout supports
only them, GSG has a significant impact. This can be shown using the S3 DFG set 
and by starting from a full layout that supports only these operations.

\helex\ is designed to allow users not to run the GSG phase of pruning
through a command line option. 
Thus, we run two versions of \helex\ using DFG set S3 on the same CGRA sizes as in 
Section~\ref{sec:arearesults}. The \textit{full} version is the complete 
\helex\ search. The \textit{noGSG} version is \helex\ without targeting 
the {\tt Arith} group and without running GSG. Table~\ref{tab:partialsearchresult} 
gives the percentage of the \textit{full} area and power reductions that 
\textit{noGSG} achieves. On average, \textit{noGSG} obtains only 81.7\% of 
\textit{full} area reduction and 87.7\% of \textit{full} power reductions. 
This attests to the impact that the GSG phase and {\tt Arith} group have.
Nonetheless, for DFGs that do not benefit from GSG, the phase can be
optionally not run to save search time.

\begin{table}[h]
    \centering
    \vspace*{0.15in}
    \input{tables/partialsearchresult}
    \caption{Percentage of area and power reductions of \textit{noGSG}}
    \label{tab:partialsearchresult}
\end{table}

\subsection{Determining the CGRA Size}

\begin{figure}
\centering
\begin{subfigure}{0.50\textwidth}
  \centering
  \includegraphics[width=0.9\textwidth]{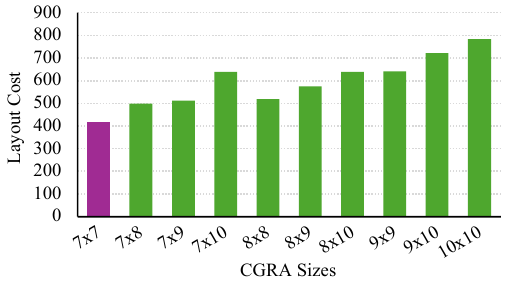}
  \vspace*{-0.05in}
  \caption{Final layout cost}
  \label{fig:cost}
\end{subfigure}
\hfill
\begin{subfigure}{0.50\textwidth}
  \centering
  \includegraphics[width=0.9\textwidth]{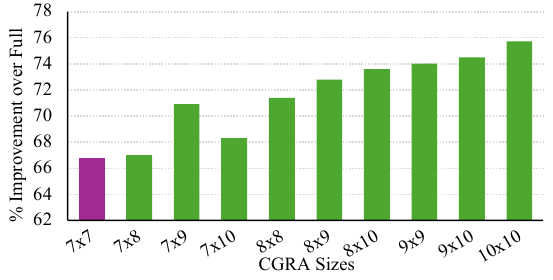}
  \vspace*{-0.05in}
  \caption{Improvement in cell cost}
  \label{fig:improv}
\end{subfigure}
\vspace*{-0.05in}
\caption{Costs and improvements in costs}
\label{fig:cgrasize}
\end{figure}

\helex\ takes as input a DFG set and a CGRA size. It is 
possible to configure \helex\ to run for a user-specified range of CGRA 
sizes and return the best CGRA size for the input DFGs. 
Fig.~\ref{fig:cgrasize} shows this for the S4 DFG set and CGRA sizes 
from 7 $\times$ 7 to 10 $\times$ 10. Fig.~\ref{fig:cost} shows the 
total cost of \helex's final layout for each size. Fig.~\ref{fig:improv} 
shows the corresponding improvement over the full layout. The smallest 
cost layout is for the 7 $\times$ 7 size, highlighted in purple. 

It is interesting to note that although the highest reduction is seen 
in the 10 $\times$ 10 CGRA, the size with the lowest cost is the 
7 $\times$ 7 one. This is because the additional cost accrued by expanding 
the CGRA outweighs the benefit of any reduction that \helex\ provides. 
For example, for the cost of increasing the 7 $\times$ 7 CGRA to a
$7 \times 8$ one requires 7 additional cells. The cost of adding these 
7 new cells without any FUs and ALUs is 66.5. In order to ``break even'',
\helex\ must remove at least an additional 66.5 cost worth of operation 
groups. This is the equivalent of approximately 11 {\tt Mult} groups, or 
4 {\tt Div} groups, which is extremely unlikely. The best CGRA size 
for any given set of DFGs is simply the {\em smallest} size that the 
set successfully maps onto, as determined by \helex.

\subsection{Impact on Latency}
\label{sec:latency}

The change in the functional layout of a CGRA may impact performance.  
DFG nodes mapped to adjacent cells in a full layout may become mapped 
to far apart cells in a heterogeneous layout, increasing latency by possibly
increasing the length of the critical path of the post-map DFG. 
However, such increase in latency does not affect steady-state
throughput since our mapper ensures a balanced DFG mappings. Thus, we 
focus on the impact of \helex\ on latency.

\begin{figure}[h]
  \centering
  \includegraphics[width=0.9\linewidth]{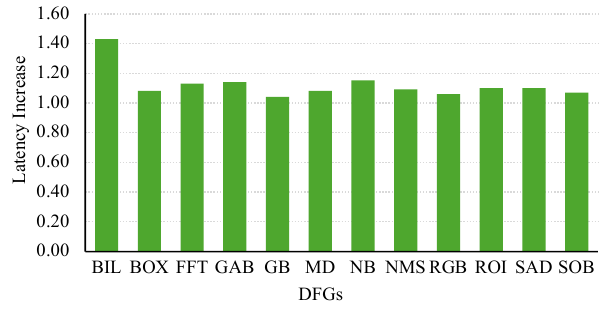}
  \vspace*{-0.10in}
  \caption{\helex's impact on latency}
  \label{fig:latency}
\end{figure}

Fig.~\ref{fig:latency} shows for each DFG the post-mapping increase
in DFG latency for the best layout relative to the full layout,
averaged over all the configurations evaluated in 
Sections~\ref{sec:arearesults}~and~\ref{sec:dfgimpact}.
The latency is measured by the length of the 
critical path of a DFG once it is mapped onto the CGRA. There is 
minimal increase in latency across the DFGs; on average 1.12X higher 
with a maximum of 1.43X. This is the increase in initial 
latency of a pipelined execution; its impact depends on 
the length of steady state execution.

BIL shows the maximum increase of 1.43X. Its DFG has several of the 
more expensive operations (FDIV and EXP) that feed one another. These 
operations are less common in other DFGs. These expensive operations 
are sometimes mapped far apart, leading to latency increase when they 
are chained, as in BIL. This suggests that it may be beneficial to 
cluster such operations together and map them as a unit, which is left 
for future work.

\subsection{Comparison to State-of-the-Art}
\label{sec:stateoftheart}

\helex's removal of cell functionality is compared to that of two 
state-of-the-art frameworks: HETA~\cite{heta} and REVAMP~\cite{revamp}. 
Since the memory and interconnect resources of the CGRAs used by HETA, 
REVAMP, and \helex\ differ greatly, the comparison is limited to 
the compute resource savings obtained under spatial configuration. 

Both frameworks do not support many operations in our 
DFGs. Extending them to support these operations is beyond 
the scope of this work. Thus, we utilize the 8 DFGs used in HETA's 
evaluation~\cite{heta}, shown in Table~\ref{tab:hetadfgs}, 
and target a $20 \times 20$ CGRA (as defined by 
each framework), which is the smallest {\em spatial} CGRA size that 
HETA successfully maps the DFGs with its default settings.
For HETA, this is a $20 \times 20$ grid of compute 
cells and two columns of 20 load store units (LSU) each. For REVAMP, this is a 
$20 \times 20$ grid of PEs that can support compute or memory operations. For 
\helex, this is an $18 \times 18$ inner grid of compute cells and 76 IO cells 
on the boundary (i.e., a $20\times 20$ in total). 

\begin{table} 
    \centering
    \input{tables/hetadfgs} 
    \caption{DFGs used in comparison, sourced from \cite{heta}}
    \label{tab:hetadfgs}
\end{table} 

HETA is run to collect reductions in the number of Add/Sub and Mult 
operations. REVAMP, on the other hand, uses a \textit{hotspot index} 
(similar to our heatmap) to determine the necessary resources 
for the functional layout~\cite{revamp}. This hotspot index based
layout remains static and is not further optimized by the framework;
only memory are interconnect resources are. This make it possible 
for us to follow the steps outlined in~\cite{revamp} and determine 
the number of operations required in the functional layout and 
calculate the resulting reduction, without the need to run REVAMP.

\begin{figure}[t]
  \centering
  \includegraphics[width=0.90\linewidth]{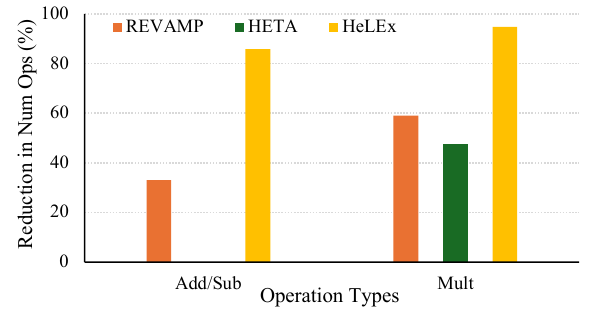}
  \vspace*{-0.1in}
  \caption{Comparison to existing work}
  \label{fig:existingwork-comparison}
\end{figure}

Fig.~\ref{fig:existingwork-comparison} shows the reduction in the number of 
PEs that support Add/Sub and Mult operations of the final heterogeneous 
layout generated by each framework, compared to the initial $20 \times 20$ 
homogeneous CGRA. \helex\ removes up to 2.6X more excess compute 
resources compared to HETA and REVAMP. It should be noted that HETA does
not report any reduction in the total number of Add/Sub operations.

For the 8 DFGs, both HETA and \helex\ report results in $\sim$7 hours. It is 
important to note that HETA also attempts to optimize network and memory 
resources during its main search time. However, the runtime for both 
\helex\ and HETA are on the same order of magnitude given the same DFGs 
and CGRA size. 

It is difficult to directly compare \helex's and REVAMP's runtime because
the latter only maps the set of DFGs once at the beginning of its search
to determine the hotspot index layout~\cite{revamp}. Thus, it is more 
appropriate to compare the time required by \helex\ to generate the heatmap layout 
against the time required by REVAMP to initially map the set of DFGs and
generate its hotspot index. 
The evaluation in~\cite{revamp} reports this mapping takes $10^3$ seconds 
on average.  \helex\ takes a comparable amount of time, roughly $10^2$ seconds.
It is important to note that there is more reduction achievable beyond the heatmap or 
\textit{hotspot index}, as shown in Section~\ref{sec:arearesults}, justifying
the additional search time by \helex.

%% file: tables/dfgs-used.tex
\setlength{\tabcolsep}{2pt}
\begin{tabular}{
    lccl
}
\toprule 
\textbf{DFG} & \textbf{Nodes} & \textbf{Edges} & \hspace*{0.25in} \textbf{Description} \\\midrule
BIL     &   26  &   29  &   Bilateral Filter Kernel                         \\ 
BOX     &   19  &   18  &   Box Filter Kernel                               \\
FFT     &   54  &   68  &   Radix-4 Fast Fourier Transform Kernel           \\ 
GAR     &   21  &   24  &   Gabor Filter Kernel                             \\ 
GB      &   16  &   12  &   Gaussian Blur Filter Kernel                     \\ 
MD      &   55  &   74  &   Molecular Dynamics Simulation Kernel            \\ 
NB      &   30  &   37  &   N-Body Simulation Kernel                        \\ 
NMS     &   29  &   36  &   Non-Maximal Suppression Kernel                  \\ 
RGB     &   27  &   30  &   RGB to YIQ Converter Kernel                     \\ 
ROI     &   45  &   56  &   Region of Interest Alignment Kernel             \\ 
SAD     &   80  &   79  &   Sum of Absolute Differences Kernel              \\ 
SOB     &   9   &   8   &   Sobel Filter Kernel                             \\ 
\bottomrule
\end{tabular}

%% file: tables/cellcost.tex
\setlength{\tabcolsep}{4pt}
\renewcommand{\arraystretch}{1.0}

\begin{tabular}{l|c|l}\toprule
\textbf{Component} &\textbf{Cost} & \textbf{\hspace*{0.25in}Description} \\\midrule
\textbf{Arith ALU}              &1.0    &ALU for Arith group \\
\textbf{FP ALU}                 &4.4    &ALU for FP group \\
\textbf{Mult ALU}               &6.2    &ALU for Mult group \\
\textbf{Div ALU}                &17.0   &ALU for Div group \\
\textbf{Other ALU}              &12.3   &ALU for Other group \\
\textbf{FIFO (4x4x32)}          &4.9    &Cell input FIFOs \\
\textbf{Empty Cell}             &4.6    &Empty cell with no FIFOs \\
\textbf{I/O Cell}               &11.9   &Complete I/O cell \\
\bottomrule
\end{tabular}

%% file: tables/searchperf.tex
\begin{threeparttable}
\begin{tabular}{l|rr|ccc}\toprule
\textbf{Size} &\textbf{$S_{exp}$} &\textbf{$S_{tst}$} &\textbf{$T_{opsg}$} &\textbf{$T_{gsg}$} &\textbf{$T_{total}$} \\\midrule
\textbf{10 x 10}\tnote{*}  &2.22e+6 &2.06e+3 & 6.7 &13.1 &19.8 \\
\textbf{10 x 12}           &3.05e+6 &2.07e+3 & 3.5 &7.4 &10.9 \\
\textbf{10 x 14}           &9.05e+4 &1.52e+3 & 5.1 &7.7 &12.8 \\
\textbf{11 x 11}           &5.24e+6 &2.30e+3 & 2.6 &11.2 &13.8 \\
\textbf{11 x 13}\tnote{*}  &9.01e+2 &2.56e+3 & 21.0 &3.9 &24.9 \\
\textbf{11 x 15}\tnote{*}  &1.03e+6 &2.82e+3 & 10.9 &15.5 &26.4 \\
\textbf{12 x 12}\tnote{*}  &1.84e+5 &2.58e+3 & 15.1 &10.8 &25.9 \\
\textbf{12 x 14}           &9.01e+5 &2.33e+3 & 5.1 &11.7 &16.8 \\
\textbf{13 x 15}\tnote{*}  &4.22e+4 &3.20e+3 & 27.0 &9.3 &36.3 \\
\bottomrule
\end{tabular}
\begin{tablenotes}
    \item[*] The initial layout for these sizes is the full layout \hfill \break
\end{tablenotes}
\end{threeparttable}

%% file: tables/dc-validation.tex


\begin{tabular}{lcc|cc|cc|c}\toprule
\textbf{} &\textbf{Synop.} &\textbf{Synop.} &\textbf{HeLEx Est.} &\textbf{HeLEx Est.} & & &\textbf{HeLEx} \\
\textbf{} &\textbf{Area} &\textbf{Power} &\textbf{Area} &\textbf{Power} &$\mathbf{\%\Delta_{Area}}$ &$\mathbf{\%\Delta_{Power}}$ &\textbf{Cost} \\\midrule
\textbf{8 $\times$ 8 Full} &2120653 &347550 &2150828 &344292 &1.4 &0.9 &2095.9 \\
\textbf{8 $\times$ 8 Hetero} &844466 &209180 &853618 &209020 &1.1 &0.1 &833.9 \\
\textbf{\% Improve} &60.2 &39.8 &60.3 &39.3 &0.1 &0.5 &60.2 \\
\midrule
\midrule
\textbf{12x12 Full} &5454735 &663690 &5505068 &657200 &5577.6 &0.9 &1.0 \\
\textbf{12x12 Hete} &1638789 &334960 &1639046 &333253 &1660.6 &0.0 &0.5 \\
\textbf{\% Improve} &70.0 &49.5 &70.2 &49.3 &70.2 &0.2 &0.2 \\
\bottomrule
\end{tabular}

%% file: tables/fifos.tex
\setlength{\tabcolsep}{8pt}
\vspace*{0.15in}
\begin{tabular}{lc|cc}\toprule
               &\textbf{Unused} &\multicolumn{2}{c}{\textbf{$\%Impr$}} \\\cmidrule{3-4}
\textbf{Sizes} &\textbf{FIFOs}  &\textbf{A} &\textbf{P}                \\\midrule
\textbf{10 x 10} &83/400  &3.2 &6.1  \\
\textbf{10 x 12} &110/480 &3.3 &6.5  \\
\textbf{10 x 14} &132/560 &3.4 &6.5  \\
\textbf{11 x 11} &115/484 &3.5 &6.7  \\
\textbf{11 x 13} &152/572 &3.8 &7.2  \\
\textbf{11 x 15} &197/660 &4.1 &7.9  \\
\textbf{12 x 12} &141/576 &3.4 &6.6  \\
\textbf{12 x 14} &198/672 &4.0 &7.8  \\
\textbf{13 x 15} &276/780 &3.7 &9.1  \\
\bottomrule
\end{tabular}

%% file: tables/dfgsets.tex
\begin{threeparttable}
\setlength{\tabcolsep}{5pt}
\renewcommand{\arraystretch}{1.10}
\begin{tabular}{c|cll|c}
\toprule 
\textbf{Set ID} & \textbf{\# of DFGs} & \textbf{DFGs} & \textbf{Description} & \textbf{Configurations} \\ \midrule
\textbf{S1} &   3 & GAR, NMS, ROI                   & Small set of DFGs        & $7 \times 9$, $9 \times 11$ \\
\textbf{S2} &   4 & BIL, NB, NMS, RGB               & DFGs of similar size     & $7 \times 7$, $9 \times 9$ \\
\textbf{S3} &   4 & FFT, GB, RGB, SOB               & Arith and Mult only DFGs & $10 \times 10$, $12 \times 12$ \\
\textbf{S4} &   5 & BIL, BOX, GB, GAR, SOB          & Image processing DFGs    & $7 \times 7$, $9 \times 9$ \\
\textbf{S5} &   6 & BIL, GB, MD, NB,                & Large set of DFGs        & $9 \times 9$, $11 \times 11$ \\
            &     & ROI, SOB                        &                          &  \\
\textbf{S6} &   7 & BIL, MD, NB, RGB,               & Large set of DFGs        & $10 \times 10$, $12 \times 12$ \\
            &     & ROI, SAD, SOB                   &                          &  \\
\bottomrule
\end{tabular}
\end{threeparttable}

%% file: tables/partialsearchresult.tex
\begin{tabular}{l|cc}\toprule
\textbf{Config} &\textbf{$Original/Partial_{area}$} &\textbf{$Original/Partial_{power}$} \\\midrule
\textbf{10$\times$10 S3} &0.83\%  &0.89\%  \\
\textbf{10$\times$12 S3} &0.81\%  &0.87\%  \\
\bottomrule
\end{tabular}

%% file: tables/hetadfgs.tex
\setlength{\tabcolsep}{8pt}

\begin{tabular}{l|r|r|c|c|c}\toprule
\textbf{DFG} &\textbf{V} &\textbf{E} &\textbf{Add/Sub} &\textbf{Mult} &\textbf{Load/Store} \\\midrule
\textbf{arf} &46 &48 &12 &16 &18 \\
\textbf{centro-fir} &46 &60 &20 &8 &18 \\
\textbf{cosine2} &82 &91 &26 &16 &40 \\
\textbf{ewf} &43 &56 &26 &8 &9 \\
\textbf{fft} &37 &48 &12 &8 &17 \\
\textbf{fir} &44 &43 &10 &11 &23 \\
\textbf{resnet2} &64 &63 &15 &16 &33 \\
\textbf{stencil3d} &66 &68 &25 &7 &34 \\
\bottomrule
\end{tabular}

%% file: related.tex
\section{Related Work}
\label{sec:related}

There are several state-of-the-art approaches that explore heterogeneity in the 
compute resources of CGRAs. They range from manual CGRA design tools~\cite{Opencgra, 
Cgrame}, to automated design space exploration (DSE) frameworks~\cite{revamp, 
hiercgra, heta, Aurora}, and customized PE design frameworks~\cite{ansaloni2010egra, 
melchert2023apex, radish, Prabhakar17}. Many of these works only target temporally 
configured, synchronous CGRAs and thus are difficult to directly compare against 
with \helex.  Thus, we review salient approaches that can target spatially configured 
CGRAs.

HETA~\cite{heta} and HierCGRA~\cite{hiercgra} utilize Bayesian optimization 
to iteratively remove excess compute, interconnect, and memory resources from 
a CGRA design given a set of input DFGs. HierCGRA introduces a hierarchical 
mapping algorithm to improve DFG mapping. In contrast to both frameworks, 
\helex\ uses a BB search-based approach to perform the removal, but it focuses 
on spatial CGRAs. Also, it performs memory optimization as a post-processing step. 
Comparison to HETA in Section~\ref{sec:stateoftheart} (the HierCGRA repository 
provided by~\cite{hiercgra} does not contain the full DSE framework), shows 
that \helex\ achieves better results. 

REVAMP~\cite{revamp} also optimizes for compute, interconnect, and memory 
resources for a target CGRA, given a set of DFGs. However, it uses a ``one-shot'' 
static approach to determine the resources necessary in the functional layout of 
a heterogeneous CGRA. The approach is based on a calculation of the frequency that 
each operation appears in the set of DFGs as well as a \textit{hotspot index}
that is generated by individual DFG mappings that determines the maximum number 
of operations each PE should contain. In this respect, our heatmap is similar, 
but \helex\ uses the heatmap as a possible starting point for its search to 
further improve it. Additionally, REVAMP utilizes Synopsys DC to synthesize each 
generated CGRA design while \helex\ only utilizes it once to generate the cost 
estimation model, avoiding lengthy synthesis time. 

\helex's approach can be viewed as complementary to that of the above frameworks.
It provides superior functional layouts, as shown in Section~\ref{sec:stateoftheart}.
These layouts can be provided as input to these frameworks to further have memory
and interconnect resource optimized, offer even more area and power savings than 
either approach can obtain independently.

APEX~\cite{melchert2023apex} uses frequent subgraph analysis to customize the
design of PEs in a CGRA, where each PE implements the functionality of an entire 
subgraph. RADISH~\cite{radish} uses a genetic algorithm to iteratively combine 
smaller PEs in larger, more complex ones. EGRA~\cite{ansaloni2010egra} introduces 
the concept of an even more coarse-grained CGRA design using reconfigurable 
ALU clusters (RACs) that implement the functionality of entire expressions 
rather than operations in a  DFG. While these works also aim to improve the 
heterogeneity of CGRAs, they ultimately solve a different problem than \helex.

%% file: conc.tex
\section{Concluding Remarks}
\label{sec:conclusion}

\helex\ is a novel framework for generating 
heterogeneous spatial and elastic CGRA functional layouts. 
Given a set of input DFGs and a target CGRA size, \helex\
utilizes a branch-and-bound search in order to eliminate
functionality of a homogeneous CGRA. It results in a 
functional layout that minimizes area and
power, while ensuring the mapping success of
the input DFGs. 

Experimental evaluation shows that the use of \helex\ 
results in layouts with significant area and power savings,
roughly around 69\% and 50\%, respectively. These savings
area realized across different CGRA sizes and across 
different combinations of input DFGs. In addition, the
use of the heterogeneous layouts generated by \helex\
has minimal impact on the latency of DFG execution.
Finally, \helex\ generates more favorable layouts compared
to those generated by two state-of-the-art frameworks. 

\helex\ focuses on the functional layout. Thus, it 
complements existing frameworks that also target other
CGRA resources. Its superior functional layouts can serve
as a starting point for these other frameworks to further
remove other resources.

There are several directions for future work. \helex\
can be extended to support temporal CGRA architectures and mappers. 
More exploration can be done for CGRA architectures with 
more complex interconnect networks and memory resources. Finally, 
analysis can be done on the impact of different operation 
groupings, particularly clustering chained operations together, 
different cost modeling approaches, and different DFG sets. 

%% file: paperbib.bib
@inproceedings{Prabhakar17,
  author={Prabhakar, Raghu and Zhang, Yaqi and Koeplinger, David and 
          Feldman, Matt and Zhao, Tian and Hadjis, Stefan and Pedram, Ardavan 
          and Kozyrakis, Christos and Olukotun, Kunle},
  booktitle={Int'l Symp. on Computer Architecture}, 
  title={Plasticine: A reconfigurable architecture for parallel patterns}, 
  year={2017},
  pages={389-402}
}

@inproceedings{Aurora,
  title={Aurora: {A}utomated refinement of coarse-grained reconfigurable accelerators},
  author={Tan, Cheng and Xie, Chenhao and Li, Ang and Barker, Kevin J 
          and Tumeo, Antonino},
  booktitle={Proc. of Design, Automation \& Test in Europe Conference \& Exhibition (DATE)},
  pages={1388--1393},
  year={2021},
}

@inproceedings{Opencgra,
  title={Opencgra: {D}emocratizing coarse-grained reconfigurable arrays},
  author={Tan, Cheng and Agostini, Nicolas Bohm and Zhang, Jeff and 
          Minutoli, Marco and Castellana, Vito Giovanni and Xie, Chenhao 
          and Geng, Tong and Li, Ang and Barker, Kevin and Tumeo, Antonino},
  booktitle={Proc. of the Int'l Conf. on Application-Specific Systems, 
             Architectures and Processors},
  pages={149--155},
  year={2021},
}

@inproceedings{Cgrame,
  title={{CGRA-ME}: {A} unified framework for {CGRA} modelling and exploration},
  author={Chin, S Alexander and Sakamoto, Noriaki and Rui, Allan and Zhao, 
          Jim and Kim, Jin Hee and Hara-Azumi, Yuko and Anderson, Jason},
  booktitle={Proc. of the Int'l Conf. on Application-Specific Systems, 
             Architectures and Processors},
  pages={184--189},
  year={2017},
}

@article{radish,
  title={Iterative search for reconfigurable accelerator blocks with a compiler in the loop},
  author={Willsey, Max and Lee, Vincent T and Cheung, Alvin and Bod{\'\i}k, Rastislav and Ceze, Luis},
  journal={Trans. on Computer-Aided Design of Integrated Circuits and Systems},
  volume={38},
  number={3},
  pages={407--418},
  year={2018},
}

@misc{freepdk45,
    author = {Christopher Torng},
    title = {FreePDK45 and the Nangate Open Cell Library},
    howpublished = {\url{https://mflowgen.readthedocs.io/en/latest/stdlib-freepdk45.html}},
    year = 2020
}

@misc{designware,
    author = {Synopsys},
    title = {DesignWare Library - Datapath and Building Block IP},
    howpublished = {\url{https://www.synopsys.com/dw/buildingblock.php}},
    year = {2024}
}

@misc{synopsys,
author = {Synopsys},
title = {Design Compiler},
howpublished={\url{https://www.synopsys.com/implementation-and-signoff/rtl-synthesis-test/dc-ultra.html}},
year = {2024}
}

@inproceedings{Ragheb22,
  author={Ragheb, Omar and Yu, Tianyi and Beidas, Rami and Anderson, Jason},
  booktitle={Proc. of the Int'l Parallel and Distributed Processing Symp. Workshops}, 
  title={Elastic Multi-Context {CGRAs}}, 
  year={2022},
  pages={655-662},
}

@inproceedings{Podobas20,
  author={Podobas, Artur and Sano, Kentaro and Matsuoka, Satoshi},
  booktitle={Proc. of the Int'l Conf. on Application-specific Systems, Architectures and Processors}, 
  title={A Template-based Framework for Exploring Coarse-Grained Reconfigurable Architectures}, 
  year={2020},
  pages={1-8},
}

@article{Podobas20survey,
  author={Podobas, Artur and Sano, Kentaro and Matsuoka, Satoshi},
  journal={IEEE Access}, 
  title={A Survey on Coarse-Grained Reconfigurable Architectures From 
         a Performance Perspective}, 
  year={2020},
  volume={8},
  number={},
  pages={146719-146743},
}

@article{Suganth09,
  author={Paul, Suganth and Jayakumar, Nikhil and Khatri, Sunil P.},
  journal={Trans. on Very Large Scale Integration (VLSI) Systems}, 
  title={A Fast Hardware Approach for Approximate, Efficient Logarithm and Antilogarithm Computations}, 
  year={2009},
  volume={17},
  number={2},
  pages={269-277},
}

@inproceedings{Nowatzki17,
  author={Nowatzki, Tony and Gangadhar, Vinay and Ardalani, Newsha and 
          Sankaralingam, Karthikeyan},
  booktitle={Int'l Symp. on Computer Architecture}, 
  title={Stream-dataflow acceleration}, 
  year={2017},
  pages={416-429}
}

@article{Morrison2016,
author = {David R. Morrison and Sheldon H. Jacobson and Jason J. Sauppe and Edward C. Sewell},
title = {Branch-and-bound algorithms: A survey of recent advances in 
         searching, branching, and pruning},
journal = {Discrete Optimization},
volume = {19},
pages = {79-102},
year = {2016},
}

@inproceedings{melchert2023apex,
  title={Apex: A framework for automated processing element design space exploration 
         using frequent subgraph analysis},
  author={Melchert, Jackson and Feng, Kathleen and Donovick, Caleb and Daly, Ross and Sharma, Ritvik and Barrett, Clark and Horowitz, Mark A and Hanrahan, Pat and Raina, Priyanka},
  booktitle={Proc. of the Int'l Conf. on Architectural Support for Programming Languages 
             and Operating Systems},
  pages={33--45},
  year={2023}
}

@inproceedings{Mei03,
author={Mei, Bingfeng and Vernalde, Serge and Verkest, Diederik and De Man, Hugo
        and Lauwereins, Rudy},
title={{ADRES}: An Architecture with Tightly Coupled VLIW Processor and 
       Coarse-Grained Reconfigurable Matrix},
booktitle={Proc. of the Int'l Conf. on Field Programmable Logic and Application},
year={2003},
pages={61-70},
}

@article{Liu2019,
author = {Liu, Leibo and Zhu, Jianfeng and Li, Zhaoshi and Lu, Yanan and Deng, 
          Yangdong and Han, Jie and Yin, Shouyi and Wei, Shaojun},
title = {A Survey of Coarse-Grained Reconfigurable Architecture and Design: Taxonomy, 
         Challenges, and Applications},
year = {2019},
volume = {52},
number = {6},
journal = {ACM Comput. Surv.},
}

@inproceedings{Lee09,
  author={Lee, Dongwook and Jo, Manhwee and Han, Kyuseung and Choi, Kiyoung},
  booktitle={Proc. of the Int'l Conf. on Field-Programmable Technology}, 
  title={{FloRA}: Coarse-grained reconfigurable architecture with 
         floating-point operation capability}, 
  year={2009},
  pages={376-379},
}

@inproceedings{Kasgen21,
  author={Käsgen, Philipp and Messelka, Mohamed and Weinhardt, Markus},
  booktitle={Proc. of the Int'l Conf. on Field-Programmable Logic and Applications}, 
  title={{HiPReP}: High-Performance Reconfigurable Processor - Architecture and Compiler}, 
  year={2021},
  pages={380-381},
}

@inproceedings{Karunaratne19,
  author={Karunaratne, Manupa and Wijerathne, Dhananjaya and 
          Mitra, Tulika and Peh, Li-Shiuan},
  booktitle={Proc. of Int'l Conf. on Computer-Aided Design}, 
  title={{4D-CGRA}: Introducing Branch Dimension to Spatio-Temporal 
          Application Mapping on {CGRAs}}, 
  year={2019},
  pages={1-8},
}

@article{Han13,
author = {Han, Kyuseung and Ahn, Junwhan and Choi, Kiyoung},
title = {Power-Efficient Predication Techniques for Acceleration of 
         Control Flow Execution on {CGRA}},
year = {2013},
volume = {10},
number = {2},
journal = {ACM Trans. Archit. Code Optim.},
}

@article{Govindaraju12,
  author={Govindaraju, Venkatraman and Ho, Chen-Han and Nowatzki, Tony 
          and Chhugani, Jatin and Satish, Nadathur and Sankaralingam, Karthikeyan 
          and Kim, Changkyu},
  journal={IEEE Micro}, 
  title={{DySER}: Unifying Functionality and Parallelism Specialization for 
         Energy-Efficient Computing}, 
  year={2012},
  volume={32},
  number={5},
  pages={38-51},
}

@article{heta,
  author={Dai, Yuan and Li, Jingyuan and Zhu, Qilong and Qiu, Yunhui and 
          Hu, Yihan and Yin, Wenbo and Wang, Lingli},
  journal={Trans. on Very Large Scale Integration (VLSI) Systems}, 
  title={{HETA}: A Heterogeneous Temporal {CGRA} Modeling and Design Space Exploration 
         via {Bayesian} Optimization}, 
  year={2024},
  volume={32},
  number={3},
  pages={505-518},
}

@inproceedings{Chin18,
  author={Chin, S. Alexander and H. Anderson, Jason},
  booktitle={Proc. of Design Automation Conf.}, 
  title={An Architecture-Agnostic Integer Linear Programming Approach to {CGRA} Mapping}, 
  year={2018},
  pages={1-6},
}

@inproceedings{Chen18,
  author={Chen, Tao and Srinath, Shreesha and Batten, Christopher and Suh, G. Edward},
  booktitle={Proc. of Int'l Symp. on Microarchitecture}, 
  title={An Architectural Framework for Accelerating Dynamic Parallel Algorithms on 
         Reconfigurable Hardware}, 
  year={2018},
  pages={55-67}
}

@article{hiercgra,
author = {Chen, Sichao and Cai, Chang and Zheng, Su and Li, Jiangnan and Zhu, Guowei 
          and Li, Jingyuan and Yan, Yazhou and Dai, Yuan and Yin, Wenbo and Wang, Lingli},
title = {{HierCGRA}: A Novel Framework for Large-scale {CGRA} with Hierarchical Modeling 
         and Automated Design Space Exploration},
year = {2024},
volume = {17},
number = {2},
journal = {ACM Trans. Reconf. Technol. Syst.},
}

@inproceedings{rodmap,
author = {Chen, Kyle Zhao Bin and Abdelrahman, Tarek S. and Azimi, 
          Reza and Czajkowski, Tomasz S. and Goudarzi, Maziar},
title = {{RoDMap}: A Reserve-on-Demand Mapper for Spatially-Configured 
         Coarse-Grained Reconfigurable Arrays},
year = {2024},
booktitle = {Proc. of the Int'l Conf. on Parallel Processing},
pages = {876-886},
}

@inproceedings{revamp,
author = {Bandara, Thilini Kaushalya and Wijerathne, Dhananjaya and 
          Mitra, Tulika and Peh, Li-Shiuan},
title = {{REVAMP}: a systematic framework for heterogeneous {CGRA} realization},
year = {2022},
booktitle = {Proc. of the Int'l Conf. on Architectural 
             Support for Programming Languages and Operating Systems},
pages = {918-932},
}

@inproceedings{Balasubramanian18,
  author={Balasubramanian, Mahesh and Dave, Shail and Shrivastava, Aviral 
          and Jeyapaul, Reiley},
  booktitle={Proc. of Design, Automation and Test in Europe Conf.}, 
  title={{LASER}: A hardware/software approach to accelerate complicated loops on {CGRAs}}, 
  year={2018},
  pages={1069-1074},
}

@article{ansaloni2010egra,
  title={{EGRA}: A coarse grained reconfigurable architectural template},
  author={Ansaloni, Giovanni and Bonzini, Paolo and Pozzi, Laura},
  journal={Trans. on Very Large Scale Integration Systems},
  volume={19},
  number={6},
  pages={1062--1074},
  year={2010},
  publisher={IEEE}
}

@inproceedings{Abbaszadeh23,
  author={Abbaszadeh, Mahdi and Abdelrahman, Tarek S. and Azimi, Reza
          and Czajkowski, Tomasz S. and Goudarzi, Maziar},
  booktitle={Proc. of the Int'l Parallel and Distributed Processing Symp. Workshops (IPDPSW)},
  title={Efficient Data Streaming for a Tightly-Coupled Coarse-Grained Reconfigurable Array},
  year={2023},
}

@inproceedings{cgra4hpc22,
  author={Podobas, Artur and Sano, Kentaro and Anderson, Jason},
  booktitle={IEEE Int'l Parallel and Distributed Processing Symp. Workshops}, 
  title={First {Int'l Workshop on Coarse-Grained Reconfigurable Architectures for 
         High-Performance Computing} ({CGRA4HPC})}, 
  year={2022},
  volume={},
  number={},
  pages={625-626},
}
